\documentclass[12pt,twoside]{article}
\usepackage{xrb2007}
\usepackage{graphicx}
\usepackage{psfig}
\usepackage{epsf}

\begin{document}

% select your session by uncommenting the appropriate line
%\session{Jets}
%\session{Jet and Black Hole Binaries}
%\session{Faint Galactic XRB Populations}
%\session{Faint XRBs and Galactic LMXBs}
%\session{Obscured XRBs and INTEGRAL Sources}
\session{ULXs}
%\session{Extragalactic Populations}
%\session{Future Missions and Surveys}
%\session{Population Synthesis}

\shortauthor{Sivakoff et~al.}
\shorttitle{Variable LMXBs in E/S0s}

\title{Variable Low-Mass X-ray Binaries in Early-Type Galaxies}
\author{
{Gregory R.~Sivakoff},\altaffilmark{1} 
{Andr\'{e}s Jord\'{a}n}, \altaffilmark{2}
{Adrienne M.~Juett}, \altaffilmark{3}
{Craig L.~Sarazin}, \altaffilmark{4}
and
{Jimmy A. Irwin}\altaffilmark{5}
}
\altaffiltext{1}{
Department of Astronomy,
The Ohio State University,
140 W. 18th Avenue, Columbus, OH 43210-1173
}
\altaffiltext{2}{
Harvard-Smithsonian Center for Astrophysics,
60 Garden Street,
MS-67, Cambridge, MA 02138
}
\altaffiltext{3}{
NASA Goddard Space Flight Center,
Greenbelt, MD 20771%;
}
\altaffiltext{4}{
Department of Astronomy,
University of Virginia,
P. O. Box 400325,
Charlottesville, VA 22904-4325
}
\altaffiltext{4}{
Department of Astronomy,
909 Dennison Building,
University of Michigan,
Ann Arbor, MI 48109-1042}

\begin{abstract}
As the {\em Chandra X-ray Observatory} mission matures, increasing numbers of
nearby galaxies are being observed multiple times, sampling the variability of
extragalactic X-ray binaries on timescales extending from seconds to years. We
present results on luminous low-mass X-ray binaries from several early-type
galaxies. We show that instantaneous LMXB luminosity functions of early-type
galaxies do not significantly change between observations; a relatively low
fraction of sources are strongly variable on $\la {\rm 5 yr}$ timescales. We
discuss the implications that a relatively small number of transient LMXBs are
being discovered in early-type galaxies.
\end{abstract}

\section{Introduction}

The detailed study of Galactic X-ray binaries (XRBs) has placed strong
constraints on theories of XRB evolution and accretion \citep[see reviews
of][]{GRS_TH2006,GRS_K2006}. These studies are limited by the $\sim 300$ active
Galactic XRBs \citep{GRS_LPH2006,GRS_LPH2007}. Fewer XRBs have well determined
distances and absorption column densities, or have high luminosities ($L_X \sim
10^{38-39} {\rm \, erg \, s}^{-1}$). The luminous systems are of interest as
they likely contain neutron stars (NSs) or black holes (BHs) radiating near or
above the Eddington limit: $L_{\rm Edd} \simeq 1.3
\times 10^{38} M_{\rm co}/M_\odot {\rm \, erg \, s}^{-1}$ for spherical accretion
of hydrogen onto a compact object of mass $M_{\rm co}$.

Low-mass X-ray binaries (LMXBs) have companions of $\la 1 M_{\odot}$ that are
typically undergoing Roche-lobe overflow. If radiating near $L_{\rm Edd}$,
the entire mass of the companion can be accreted in $\la 10^{8} {\rm \, yr}$
for accretion duty cycles of unity. This would imply that observed LMXBs
formed recently. Alternatively, the duty cycle can be lower if the episodes of
near Eddington accretion are transient, and observed LMXBs need not have formed
recently. The transient behavior of LMXBs has been attributed to a disk
instability first identified in cataclysmic variables \citep{GRS_P1996}. In
short-period systems ($P_{\rm orb} \la 12 {\rm \, hr}$) where the companion is a
main-sequence star, theory predicts most LMXBs are persistent
\citep{GRS_K2006}. An evolved companion, which is somewhat surprising at these
periods, could explain the short-period transients with BHs (and some NSs)
observed to have outbursts with timescales of $\sim 10$s of days. In
longer-period binaries, transient systems are expected, and should have longer
outbursts due to their larger disks.
One such source, GRS$+$1915+105, has been in outburst for $15 {\rm \, yr}$. 
Thus, the luminosity functions of entire populations of LMXBs depend on the mass
spectrum of accretors, transient duty cycle, companion types (main-sequence, red
giant, or white dwarf), and stellar age
\citep[e.g.,][and Kalogera et~al.\ 2008 in these proceedings]{GRS_IK2006}.

With the launch of the {\em Chandra X-ray Observatory}, luminous XRBs in more
distant galaxies are now routinely being studied \citep[e.g., references
in][]{GRS_FW2006}. Extragalactic studies are complementary to Galactic studies. 
Large numbers of XRBs that share a relatively common distance and absorption
column density are revealed; however, many binary properties (e.g., compact
object mass, donor type, orbital period) cannot be directly observed. For the
old stellar populations in early-type galaxies, the X-ray binaries are LMXBs. 
Although the studies of resolved LMXBs in early-type galaxies are typically
limited to $L_X \ga 10^{37}$--$10^{38}{\rm \, erg \, s}^{-1}$, tens to hundreds
of sources can be detected in a single galaxy. X-ray variability can help shed
light on the nature of these luminous LMXBs.

As the {\em Chandra} mission matures, more early-type galaxies are being
observed multiple times, revealing LMXB variability on timescales of seconds to
years. We discuss some results from multi-epoch observations of NGC~4697,
NGC~4365, and Centaurus A (Cen A) in this paper%
\footnote{Multi-epoch observations of NGC~3397 are discussed by Brassington
et~al.\ (2008) in these proceedings.}.
For both NGC~4697 \citep[$11 {\rm \, Mpc}$;][]{GRS_JCB+2005} and NGC~4365
\citep[$20 {\rm \, Mpc}$;][]{GRS_TDB+2001}, four $\sim 40 {\rm \, ks}$ ACIS-S
observations were made $\sim 4 {\rm \, yr}$ after initial $\sim 40 {\rm \, ks}$
observations \citep*{GRS_SIB2000,GRS_SIB2001,GRS_SSI2003}. 
Recently six $\sim 100 {\rm \, ks}$ ACIS-I observations of Cen A \citep[$3.7
{\rm
\, Mpc}$, averaging 5 distance indicators from \S~6 in][]{GRS_FMS+2007} 
were analyzed with shorter archival observations taken $\sim 4$--$7 {\rm
\, yr}$ earlier. In this paper, we focus on the variability
between observations.

\section{LMXB Luminosity Functions in NGC~4697 and NGC~4365}

\begin{figure}[!t]
{\centering \leavevmode
  \epsfxsize=.45\textwidth \epsfbox{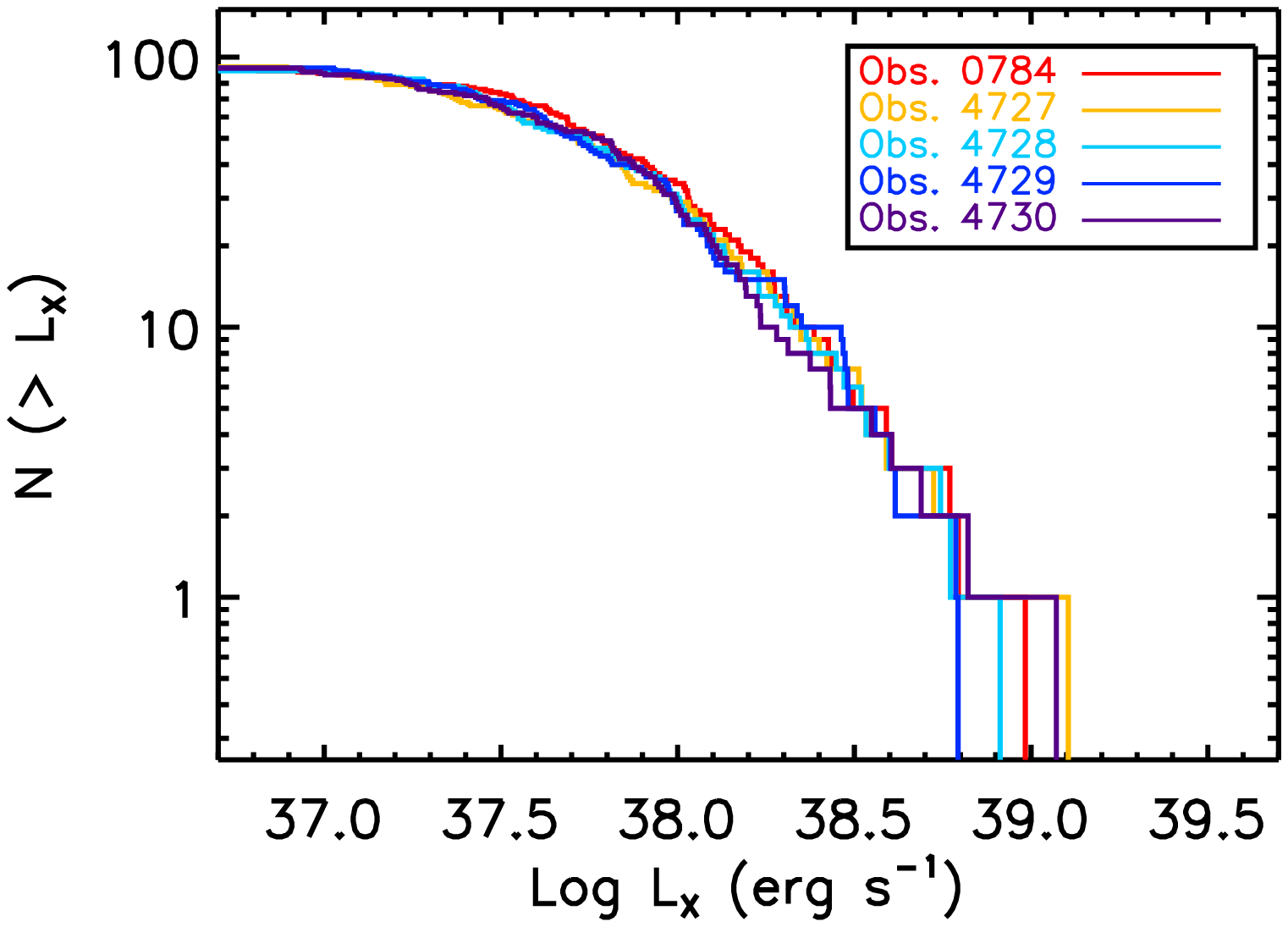} \hfil
  \epsfxsize=.45\textwidth \epsfbox{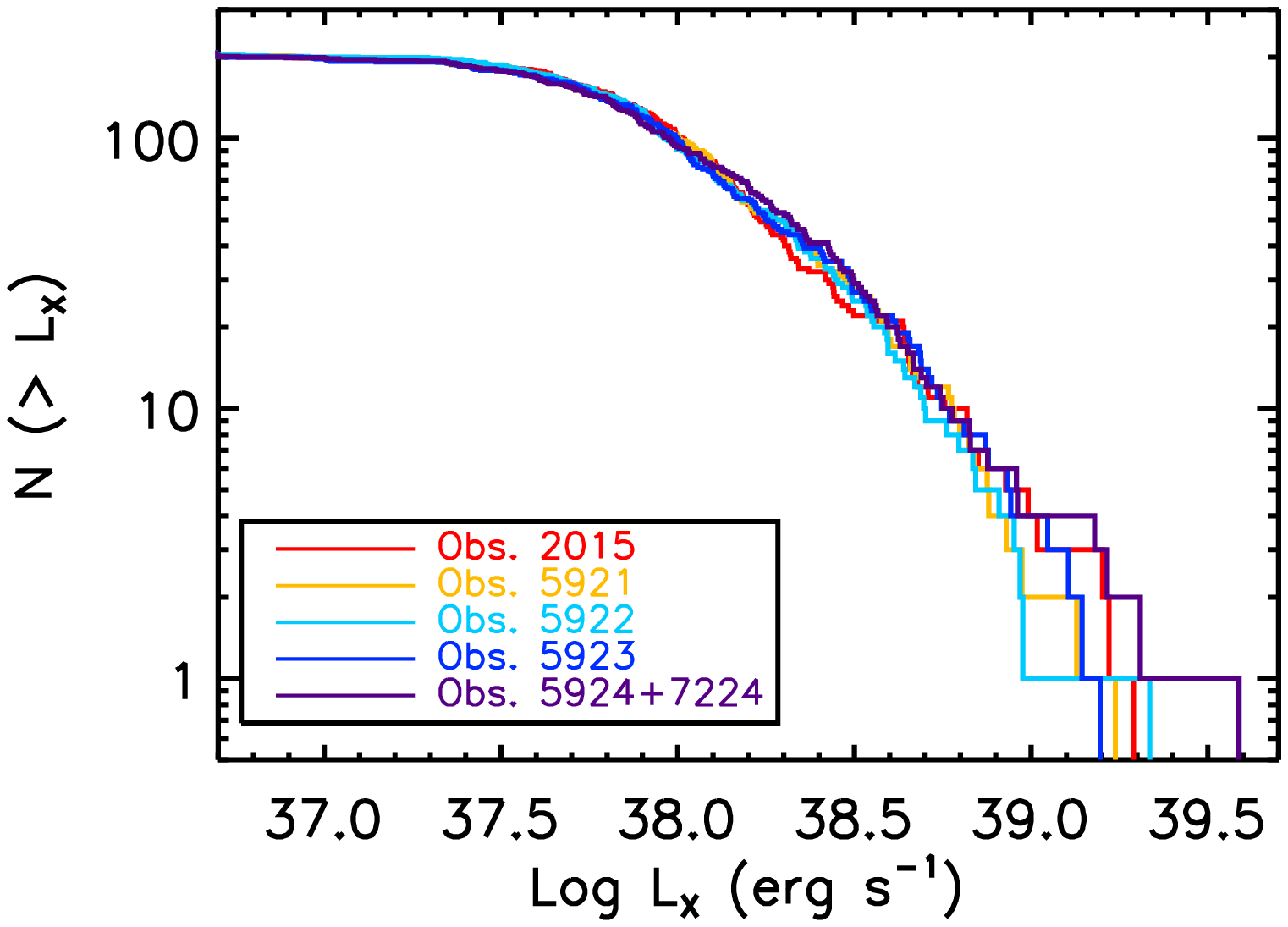}
}
\caption{
Cumulative, instantaneous luminosity functions from five
observations of NGC~4697 ({\it left}) and  NGC~4365 ({\it right}).
These functions do not vary strongly  over
$4.6 {\rm \, yr}$.
\label{grs_fig:lf}
}
\end{figure}

For NGC~4697 and NGC~4365, we performed standard reduction and flare removal. We
used {\sc CIAO wavdetect} to detect sources (158 in NGC~4697; 322 in NGC~4365)
in the combined $185 {\rm \, ks}$ (NGC~4697) and $194 {\rm \, ks}$ (NGC~4365)
observations. For each observation, we determined the count rates from
PSF-scaled source regions and local backgrounds. We converted count
rates to luminosities assuming a $9.1 {\rm keV}$ bremsstrahlung spectrum and
correcting for vignetting, the PSF, and QE degradation. We display the
cumulative instantaneous luminosity function of each observation in
Figure~\ref{grs_fig:lf}. We find no evidence for changes in the instantaneous
luminosity functions over $< 4.6 {\rm \, yr}$. 
Thus, variability on such timescales will not affect interpretations of luminosity
functions from single observations.

\begin{figure}[t]
{\centering \leavevmode
  \epsfxsize=.45\textwidth \epsfbox{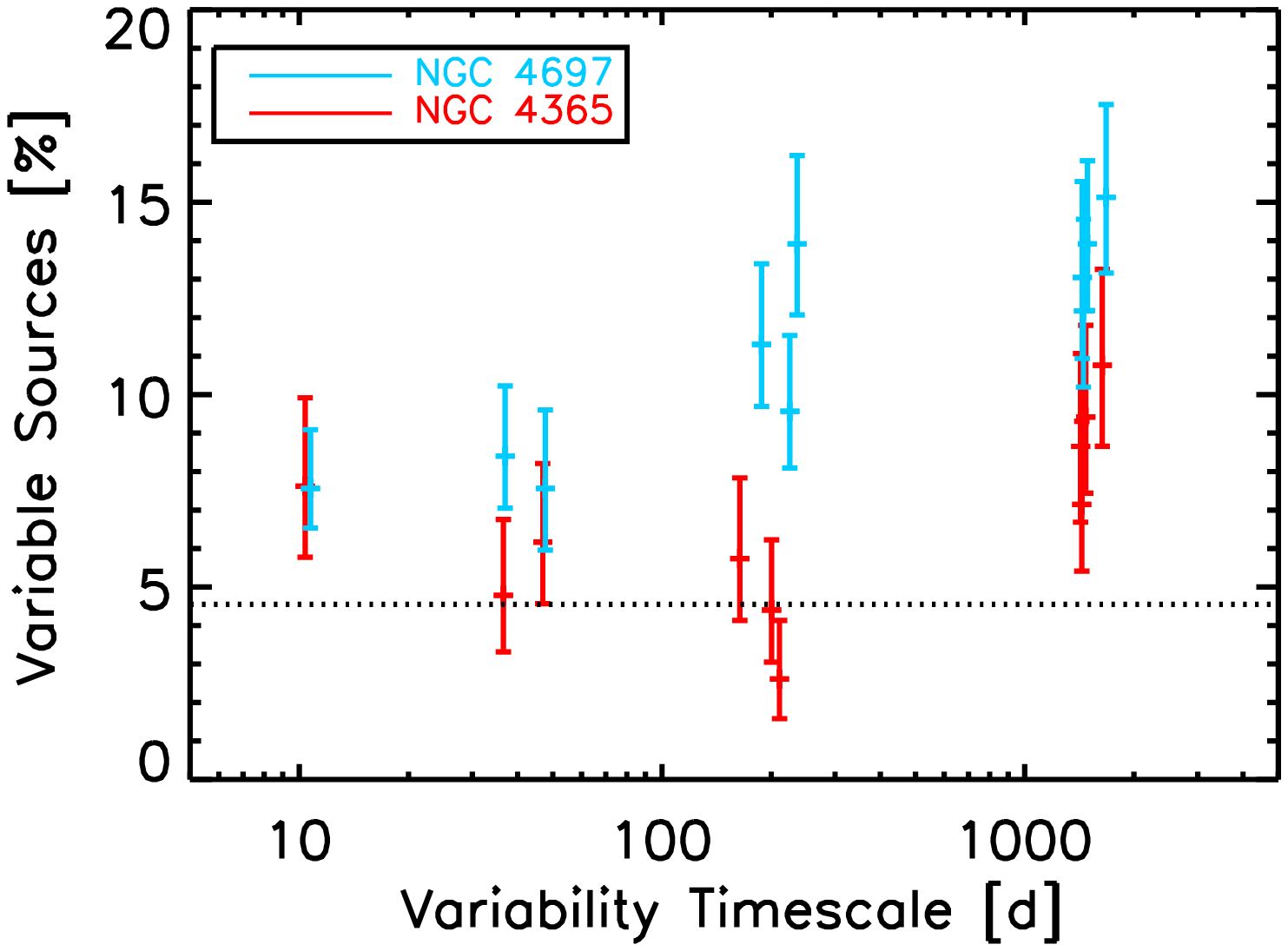} \hfil
  \epsfxsize=.225\textwidth \epsfbox{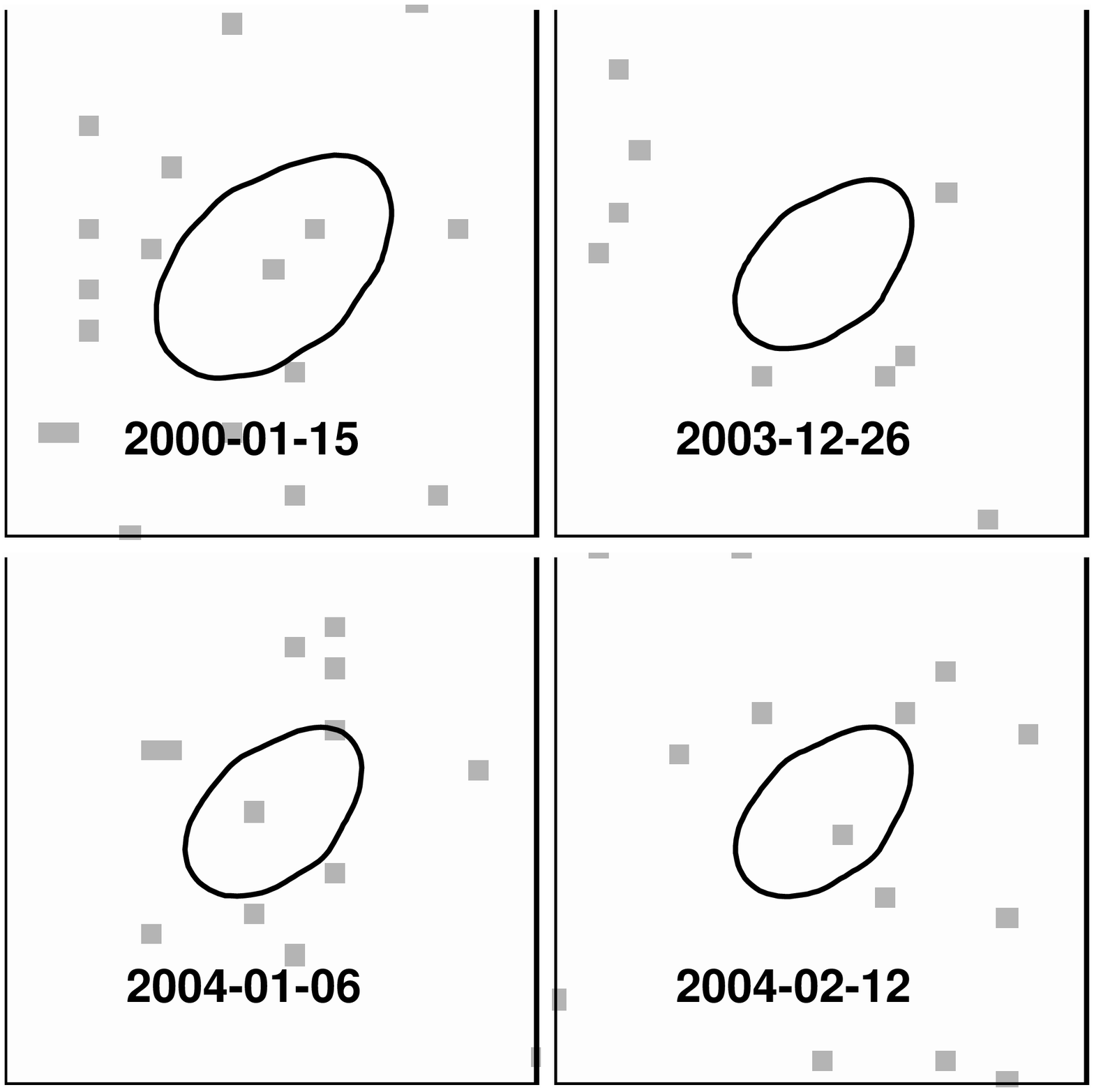}
  \epsfxsize=.225\textwidth \epsfbox{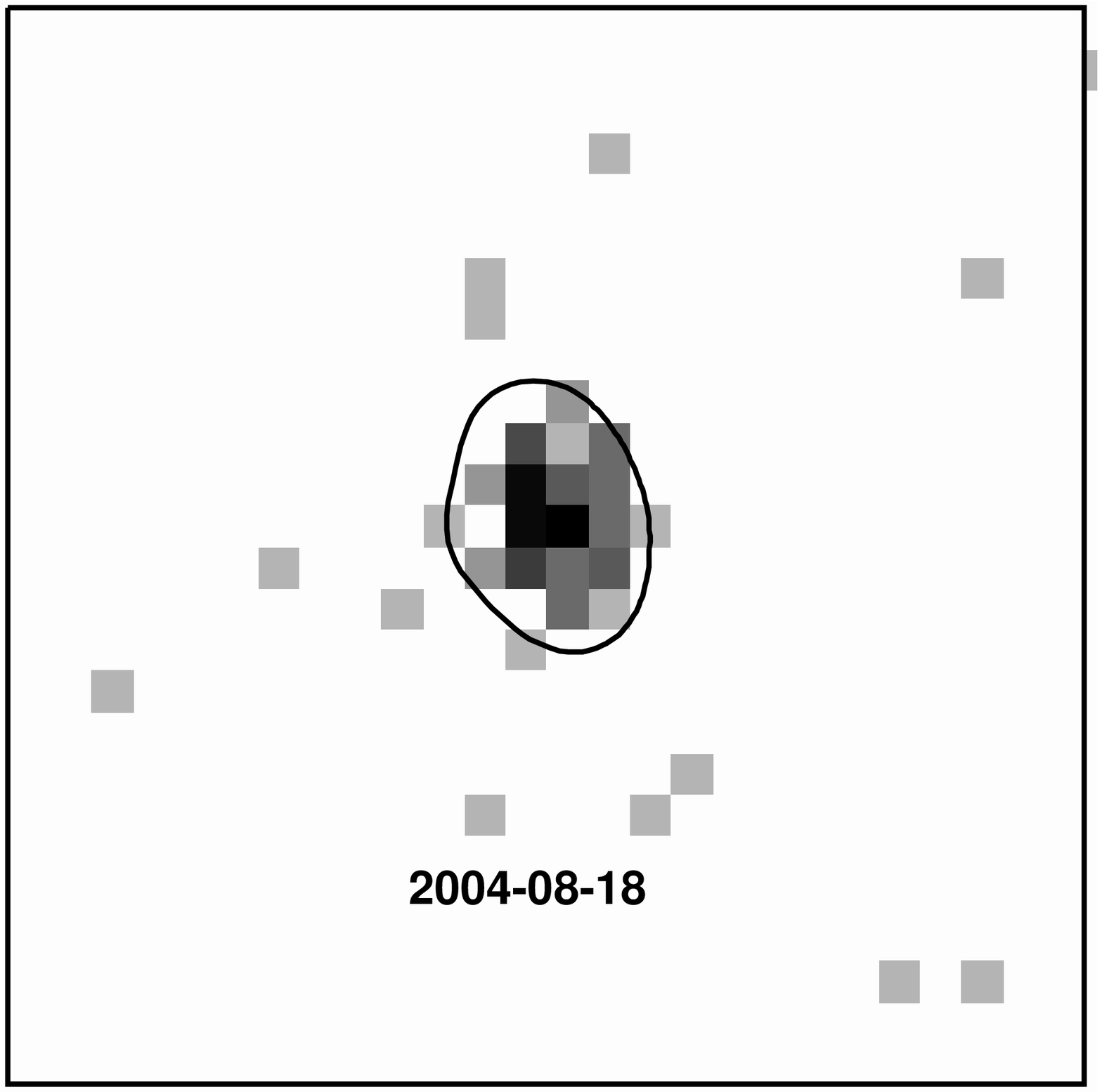}
}
\caption{
({\it Left}) The fraction of sources in NGC~4697 and NGC~4365, that are strongly
variable between any two observations as a function of the timescale between the
observations. Strongly variable LMXBs are not common. ({\it Right}) A transient
source in NGC~4697 with 87 counts in the last observation, but only 5 in all of
the previous observations.
\label{grs_fig:2pt_tr}
}
\end{figure}

The relatively constant luminosity function could result from either sources
that are not strongly variable, or sources that are strongly variable, but whose
average luminosity function is relatively constant. To address this, we
determined the fraction of sources that are strongly variable between any two
observations ({\it left} of Figure~\ref{grs_fig:2pt_tr}) or over the entire set
of observations. We classify as source as strongly variable if $\chi^2$ testing
against a constant luminosity indicates the probability a source is variable is
$>95.4\%$ (i.e, $>2\sigma$). For example, we require $\chi^2>4$ when comparing
any two observations. We only test sources whose average luminosity over all
five observation is more than $3\sigma$ above zero in calculating the fractions. 
There are 124 and 232 such sources in NGC~4697 and NGC~4365; however, the number
of these sources in a given observation is smaller due to varying
field-of-views. The fraction of strongly variable sources between any two
observations ranges from $\sim3$--$15\%$, but is roughly consistent with $8\%$. 
The fraction variable considering all five observations is $\approx16\pm4\%$ for
NGC~4697 and $\approx13\pm3\%$ for NGC~4365. Strongly variable LMXBs are not
common.

\section{Transient LMXBs}

Most of the sources in NGC~4697 and NGC~4365 are persistent over $4.6 {\rm
\, yr}$ timescales. This leads to two potential interpretations: the
majority of observed sources are persistent LMXBs or are transients sources with
outbursts that last $\gg 5 {\rm \, yr}$.

Among the strongly variable LMXBs, we identify several sources as transient
candidates. We do this by grouping each variable source into two luminosity
states. Transient candidates are sources where the luminosity of the fainter
(quiescent) state is less than $3\sigma$ above zero and the luminosity of the
brighter (outburst) state is more than $3\sigma$ above zero and occurs over
consecutive observations. There are eleven transient candidates in NGC~4697 and
twelve transient candidates in NGC~4365, most of which either turn on or turn
off. For each galaxy, we estimate the average outburst timescale assuming all
sources in a galaxy are transients undergoing long-duration outbursts and we
detected the beginning or end of that outburst in all of our transient
candidates. We estimate relatively consistent outburst timescales of
$\sim 100 {\rm \, yr}$ in NGC~4697 and $\sim180 {\rm \, yr}$ in NGC~4365;
however, we note that either assumption could be violated. The presence of
persistent NS-LMXBs cannot be ruled out at the $\ga 10^{37}
{\rm \, erg \, s}^{-1}$ luminosities we probe. Only four transient candidates
had an outburst luminosity more than ten times the limit on the quiescent
luminosity (at $>3\sigma$ confidence), and are considered clear transient
sources (e.g., the clear transient in NGC~4697 is displayed in the {\it right}
of Figure~\ref{grs_fig:2pt_tr}).

To mitigate these uncertainties, one can only consider the brightest LMXB
candidates ($L_X > 8 \times 10^{38} {\rm \, erg \, s}^{-1}$). Such sources will not
be weakly varying LMXBs near the detections limits and would be super-Eddington
for a typical NS-LMXB. Although multiple NS-LMXBs in
extragalactic GCs \citep*{GRS_SJS+2007,GRS_KMZ2007} could be more luminous,
multiple NSs near the Eddington limit would be required. In
\citet{GRS_Ir2006}, the lack of transient sources among luminous LMXBs was used
to imply outburst timescales of $\ga 50 {\rm \, yr}$. In Table~\ref{grs_tbl:bt},
we summarize the data for these galaxies, adding NGC~4697 and NGC~4365. 
Averaging all galaxies together, we estimate outburst timescales of $\sim 200
{\rm \, yr}$ for the brightest LMXBs.

\begin{table}[!t]
\caption{Breakdown of Luminous ($L_X > 8 \times 10^{38} {\rm \, erg s}^{-1}$) LMXBs
\label{grs_tbl:bt}}
\smallskip
{\small
\begin{tabular}{lcccc}
\tableline
\noalign{\smallskip}
Galaxy &
\ Transient Srcs. &
\ Persistent Src. &
\ Foreground/Background Srcs. &
Max Baseline\\
&
[Number]&
[Number]&
[Number]&
[yr]\\
\noalign{\smallskip}
\tableline
\noalign{\smallskip}
NGC~1399 & \phantom{$^{\rm a}$}0\phantom{$^{\rm a}$} & 21           & 3 & \phantom{$\sim$}3.3\phantom{$\sim$}\\ 
M87      & \phantom{$^{\rm a}$}0\phantom{$^{\rm a}$} & 16           & 1 & \phantom{$\sim$}5.6\phantom{$\sim$}\\
NGC~4697 & \phantom{$^{\rm a}$}1$^{\rm a}$           & \phantom{0}4 & 1 & \phantom{$\sim$}4.6\phantom{$\sim$}\\
NGC~4365 & \phantom{$^{\rm a}$}1$^{\rm b}$           & \phantom{0}9 & 1 & \phantom{$\sim$}4.6\phantom{$\sim$}\\
Total    & \phantom{$^{\rm a}$}2\phantom{$^{\rm a}$} & 50           & 6 & $\sim$4.4\phantom{$\sim$}\\  
\noalign{\smallskip}
\tableline
\end{tabular}
}
$^{\rm a}$Supersoft ULX of unknown origin;
$^{\rm b}$Candidate BH-LMXB in a globular cluster.
\end{table}

\section{Discussion}

Multi-epoch observations of LMXBs in early-type galaxies have revealed long-term
variable LMXBs; however, they are a minority ($\sim 10$--$15\%$) and do not
strongly affect the luminosity functions of LMXBs. Most LMXBs are persistent
over $\sim 5 {\rm \, yr}$ timescales, implying either a prevalence of inherently
persistent sources or that the sources are transient on longer timescales. We
argue that having all sources undergo outbursts with timescales of $\sim
100$--$200 {\rm \, yr}$ is consistent with our observations. This is
particularly true for the most luminous LMXBs. A picture is emerging where these
sources are analogs of GRS$+$1915+105, i.e., long-period binaries with an
evolved companion. Recent modeling of GRS$+$1915+105 (Wynn et al.\ 2008, in
these proceedings) is also consistent with an outburst timescale of $\sim 100
{\rm \, yr}$.

To estimate the duty cycle, one needs both an outburst duration and a recurrence
timescale. For $\sim 100 {\rm \, yr}$ outbursts, the recurrence timescale will be
beyond the reach of direct measurements, requiring alternative estimates of the
recurrence timescale. One estimate of $\sim 1500 {\rm \, yr}$ comes from
the crustal heating recurrence timescale for the Galactic NS-LMXB KS~1731$-$260
\citep{GRS_RBB+2002}. This would imply a duty cycle of $\sim 10\%$. Comparing
the number of active and quiescent LMXBs in Galactic GCs \citep{GRS_HGL+2003}
yields a similar duty cycle of $\sim 12\%$. Such duty cycles also have support
from the theoretical side. The duty cycle of GRS$+$1915+105 is estimated to be
$\sim 4\%$ (Wynn et al.\ 2008, in these proceedings) and duty cycles of $\sim
10\%$ are required to fit the luminosity functions of early-type galaxies
(Kalogera et al.\ 2008, in these proceedings). This suggests a healthy outlook
for future probes of the duty cycle by both observations and theory.

An outstanding issue is the lack of shorter-term transients, like those of
Galactic short-period transient LMXBs, detected in early-type galaxies. Only
NGC~4365 has three transients whose outburst timescale could be $\la {\rm yrs}$;
however, the observations are too shallow to confirm their transient nature. 
In the more nearby Cen A, there are two clear, luminous transients, but no
luminous persistent sources \citep{GRS_SKJ+2007}. One source is a recurring
transient with outbursts that lasts $\la 2 {\rm \, yr}$, while the single
outburst of the other has lasted $> 70{\rm \, d}$. Unless the latter source
proves to have a much longer duration outburst, the rates and types of luminous
transients in Cen A may be anomalous compared to other early-type galaxies. 
Before asserting that shorter timescale transient binaries do not exist in large
numbers in early-type galaxies, more work needs to be spent on understanding
whether the observational strategies employed were sufficient to detect such
transients. Future cadence times between observations and lengths of
individual observations may need adjustment.

\acknowledgments
I would like to thank the HST-ACS Virgo Cluster Survey Team and the Centaurus A
Very Large Project Team. Support for this work was provided by NASA through {\it
Chandra} Award Numbers GO4-5093X, AR4-5008X, and GO5-6086X, GO6-7091X,
GO7-8078X, and GO7-8105X, through HST Award Numbers HST-GO-10003.01-A,
HST-GO-10582.02-A, HST-GO-10597.03-A, and HST-GO-10835.01-A by an ARCS
fellowship, and by the F. H. Levinson Fund. I would like to thank the conference
organizers for their hard work in making this stimulating conference a success.

%\bibliographystyle{apj}
%\bibliography{xrb}

\end{document}